# Super-radiant emission from quantum dots in a nanophotonic waveguide

*Je-Hyung Kim,*[*,†,‡] *Shahriar Aghaeimeibodi,*[‡] *Christopher J. K. Richardson,*[§] *Richard P. Leavitt,*[§] *and Edo Waks*[*,‡,‖]

[†]Department of Physics, Ulsan National Institute of Science and Technology (UNIST), Ulsan 44919, Republic of Korea,

[‡]Department of Electrical and Computer Engineering and Institute for Research in Electronics and Applied Physics, University of Maryland, College Park, Maryland 20742, United States

[§]Laboratory for Physical Sciences, University of Maryland, College Park, Maryland 20740, United States

[‖]Joint Quantum Institute, University of Maryland  and the National Institute of Standards and Technology, College Park, Maryland 20742, United States

ABSTRACT.

**Future scalable photonic quantum information processing relies on the ability of integrating multiple interacting quantum emitters into a single chip. Quantum dots provide ideal on-chip quantum light sources. However, achieving quantum interaction between multiple quantum dots on-a-chip is a challenging task due to the randomness in their frequency and position, requiring local tuning technique and long-range quantum interaction. Here, we demonstrate quantum interactions between distant two quantum dots on a nanophotonic waveguide. We achieve a photon-mediated long-range interaction by integrating the quantum dots to the same optical mode of a nanophotonic waveguide and overcome spectral mismatch by incorporating on-chip thermal tuners. We observe their quantum interactions of the form of super-radiant emission, where the two dots collectively emit faster than each dot individually. Creating super-radiant emission from integrated quantum emitters could enable compact chip-integrated photonic structures that exhibit long-range quantum interactions. Therefore, these results represent a major step towards establishing photonic quantum information processors composed of multiple interacting quantum emitters on a semiconductor chip.**

The desire to build quantum photonic circuits that can implement complex quantum tasks has spurred a significant effort to integrate multiple quantum light sources on-a-chip. Quantum dots are ideal building blocks for quantum photonic circuits. When coupled to nanophotonic devices, these solid-state quantum emitters generate strong light-matter interactions that produce efficient single photon sources,[1-3] nonlinearities at the single photon level,[4] and quantum switches.[5] The majority of work to-date focused on devices composed of single quantum dots. But advanced quantum photonic circuits require multiple interacting quantum dots integrated into a single chip.[6,7] One way to achieve these interactions is through a photonic channel. Such photon-mediated interactions can act over long distances[8-10] and can be extremely efficient in nanophotonic cavities and waveguides.[2,11]

One example of photon-mediated interactions is super-radiance. When multiple emitters simultaneously couple to the same mode, they collectively emit faster than individual emitter, enabling a broad range of quantum optical interactions that generate non-classical states of light,[12,13] super-radiant lasers,[14] and entanglement between quantum bits.[15-17] A number of quantum systems such as integrated atoms,[10,15,18] superconducting circuits,[16] and atomic defects[17] have demonstrated super-radiant behavior. However, the realization of super-radiance in semiconductor systems such as quantum dots is significantly more challenging because these emitters suffer from a large inhomogeneous broadening of the emission frequency. Therefore, the difficulty of achieving multiple, identical quantum dots in a single optical mode has prevented the realization of super-radiance from independent quantum dots to-date.

In this letter we report an experimental demonstration of super-radiant emission of two quantum dots coupled to a nanophotonic waveguide. We utilize InAs/InP quantum dots coupled to photonic crystal waveguides that exhibit slow group velocities and high coupling efficiencies, and thus result in enhanced light-matter interactions and efficient coupling of single photons from quantum dots to a waveguide mode.[19-21] In order to compensate for the different resonant frequencies of the quantum dots, we incorporate built-in optical heaters that enable us to tune the resonances of the individual dots. We perform photon-correlation measurements on the output mode of the waveguide and observe a clear signature of super-radiance which manifests itself in as a bunching peak. Our approach can naturally extend to a larger number of quantum dots, enabling a scalable and controllable architecture to study long-range interacting quantum systems on a solid-state chip.

Figure 1a illustrates the level structure for two quantum dots that emit distinguishable photons. If the quantum dots are initially excited to the state $|ee\rangle$, then the system will decay into a statistical mixture of $|eg\rangle$ and $|ge\rangle$. The system will then decay to the ground state with the spontaneous emission rate of individual emitters, $\Gamma_{\mathrm{sp}}$. Figure 1b shows the case where the two emitters are identical and thus emit indistinguishable photons. In this case, the excited state $|ee\rangle$ will radiatively decays to the bright Dicke state,[16,22] $|B\rangle = (|eg\rangle + e^{ikL}|ge\rangle)/\sqrt{2}$, where $L$ is the distance between the two emitters and $k$ is the waveguide propagation constant. This state an entangled superposition of the two emitters, which produces super-radiance where the system emits into the ground state $|gg\rangle$, with a rate of $2\Gamma_{\mathrm{sp}}$, which is twice as fast as the case of distinguishable emitters. $|D\rangle = (|eg\rangle - e^{ikL}|ge\rangle)/\sqrt{2}$ represents the dark state.

Super-radiance requires multiple identical emitters coupled with a photonic mode. In our device, this mode corresponds to a guided mode in a photonic crystal waveguide. Such waveguides enhance light-matter interactions and efficiently couple photons from the emitters to the waveguide modes, particularly near the band edge where the group velocity is slow[19,20] Figure 1c shows a scanning electron microscope image of the fabricated device, composed of an InP photonic crystal with InAs quantum dots grown in the center of the membrane. (See Method for details on design and fabrication). In order to compensate for inhomogeneity in the quantum dot emission, we introduce a local temperature tuning method. First, we thermally isolate the entire structure from the substrate using thin 500 nm tethers and thermally separate the waveguide into two segments by introducing a narrow gap at the middle of the waveguide, which allows us to heat each segment individually. Figure 1d shows a simulated heat propagation in thermally separated two regions on the waveguide device, FEMLAB (Comsol, Inc.). In the experiment, we fabricate a heating pad on each thermally isolated segment. By shining a laser onto the pad, we can locally tune the temperature of the quantum dots within the segment with relatively small influence on the other segment.[23,24] To extract the emissions from the waveguide, we add grating out-couplers on the ends of the waveguide that exhibit a 19% extraction efficiency from the out-coupler with an objective lens (NA=0.7) (see Methods).

A line defect in the photonic crystal structures forms waveguide modes below a light line within the TE-like photonic band gap (See Figure S1 in Supporting Information). At large $k$ values of these modes, the dispersion curves become flat, resulting in a divergence of photonic density of states. Quantum emitters coupled to this region can exhibit Purcell enhancement as well as high coupling efficiency.[25] These features make photonic crystal waveguides ideally suited for studying quantum dot super-radiance.

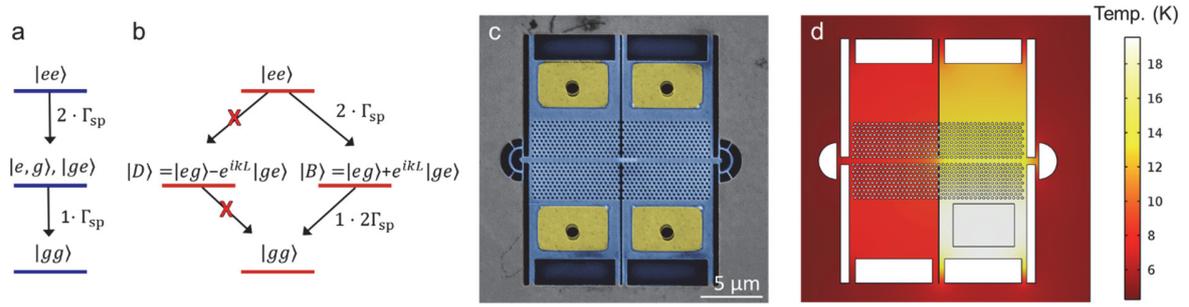

**Figure 1.** (a,b) Level structures of two two-level quantum systems for non-identical (a) and identical quantum emitters (b). $|e\rangle$ and $|g\rangle$ represent the excited and ground states of quantum dots. $|D\rangle$ and $|B\rangle$ represent the dark and bright states of coupled two quantum dots. $\Gamma_{SP}$ is the spontaneous emission rate of single quantum dots. (c) False color scanning electron microscope image of the photonic crystal waveguide device. Blue- and yellow-colored regions represent the photonic crystal devices and metal heating pads. The holes in the metal pads are formed for better undercut during the wet etching process. (d) Simulated heat transfer in the devices from FEMLAB (Comsol, Inc.). A rectangle box on the right bottom of the device represents a metal pad where the heat starts.

We characterize the device through photoluminescence measurements by exciting the center of the waveguide using a 780 nm continuous wave laser (See Figure S2 in Supporting Information). Near the band edge we measure a group index as high as 12, with a Purcell factor as large as 13. Super-radiant emission occurs only when two quantum emitters have the same resonance frequency. We first identify one quantum dot in the left region of the device and then scan the right region to find a second quantum dot with a similar wavelength. Figure 2a and 2b show photoluminescence spectra from two quantum dots we identify using this procedure. We label the quantum dot on the left as dot A and the quantum dot on the right as dot B. The figures show the spectra when we excite the dots individually, where dots A and B show an emission at 1314.5 nm and 1314.3 nm respectively, which corresponds to a spectral mismatch of 145 μeV. We select the quantum dots that are slightly detuned from the band edge of 1335 nm where the waveguide has the maximum Purcell effect but exhibits high losses due to scatter[26] and Anderson localization.[27] At chosen wavelengths, the quantum dots exhibit a moderate Purcell factor of 2~3, but they still have a high coupling efficiency of approximately 80% (See Figure S2 in Supporting Information). To confirm that these emissions originate from single quantum dots, we perform a second-order correlation measurements on each emitter. Figure 2c and 2d show the measured second-order correlation, denotes $g^{(2)}(\tau)$. Both curves show a suppressed peak at the center ($\Delta\tau = 0$), indicating single photon emission. The solid curve in each figure is a fit to a two-sided exponential function. After subtracting the detector dark counts, which we measure independently, we obtain the values of $g^{(2)}(0)$= 0.13±0.04 and 0.06±0.05 for dot A and dot B, respectively. We note that our quantum emitters generate single photons at telecom wavelength, which is important for quantum information processing on silicon chips[28] and fiber networks.

Changing the sample temperature is an effective way to tune the quantum dot wavelength. By increasing the temperature from 4 K to 26 K, we can shift the quantum dots by up to 610 μeV without a large reduction of the emission intensity or linewidth (See Figure S3 in Supporting Information). In order to locally shift the emission wavelength of dot B to the wavelength of dot A, we perform local temperature tuning by shining a 780 nm laser on a metal heat pad.[23,24] Thermally well-isolated structures allow efficient heating of one side of the waveguide without significantly modifying the temperature on the other side. Figure 2e shows the emission spectrum of dots A and B from the output of the grating coupler as a function of the heating laser power on the heat pad for dot B. As we increase the optical heating power, the wavelength of dot B shifts toward longer wavelength. The heating laser has a minimal effect on dot A, indicating that we achieve good thermal isolation. At a heating laser power of 200 μW, both dots emit at identical resonant frequencies. By comparing the wavelength shift attained from the local laser heating to that from the stage heating, we determine that dot B has a local temperature of about 15K when it resonates with dot A (See Figure S3 in Supporting Information). Figure 2f plots the emission spectra of dots A and B after local tuning. The spectra show a good spectral overlap within 10 μeV, determined by Lorentzian fitting.

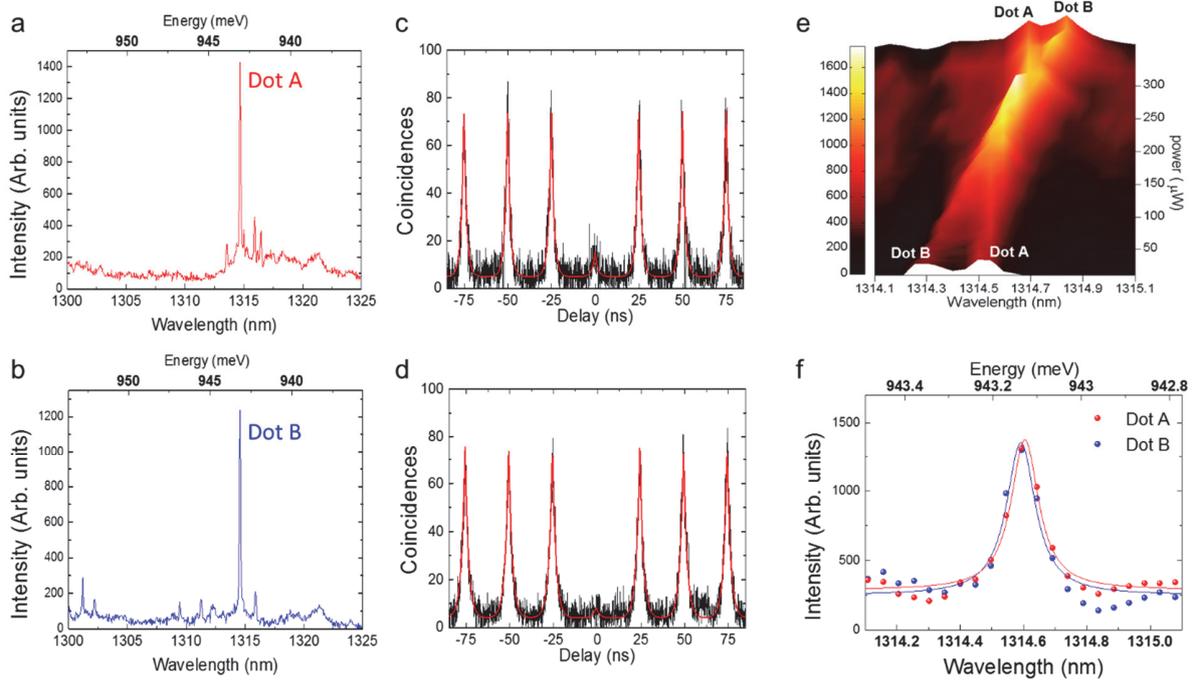

**Figure 2.** (a,b) Photoluminescence spectra of two separated dots A (a) and dot B (b) on a waveguide. (c,d) Pulsed second-order correlation histograms for dots A (c) and dot B (d). Red lines indicate fitted curves. (e) Photoluminescence spectrum plot of dots A and B as a function of heating laser power showing the crossing of the emission wavelength of dots A and B. (f) Spectrally matched two dots A and B as a result of local heating. Solid curves are fitted curves with a Lorentzian function.

We investigate the super-radiant emission by performing the second-order correlation measurements. When the two emitters are distinguishable, the second-order correlation will yield a value of $g^{(2)}(0) = 0.5$. But if the emitters are indistinguishable, the enhanced decay rate ($\Gamma'_{sp} = 2\Gamma_{sp}$) due to the super-radiance increases the probability of detecting the second photon twice resulting in a second-order correlation given by $g^{(2)}(0) = \frac{\Gamma_{sp'}}{2\Gamma_{sp}} = 1$ (See Figure S4 in Supporting Information).[17] Thus, the second-order correlation measurement provides a convenient way to probe super-radiant emission.

Figure 3a shows a setup for the second-order correlation measurements. We simultaneously excite the dots A and B and collect the photons from one grating outcoupler followed by a spectrometer, a 50:50 beamsplitter, and single photon detectors. Figure 3b shows the measured second-order correlation curves with thermal tuning (red dots) and without thermal tuning (blue dots) after subtraction of detector dark counts. The solid-lines represent a fit of both data sets to theoretical models. From the fit, in the absence of thermal tuning, the second-order correlation exhibits a dip with a width of 1.2 ns, which is consistent with the radiative decay rate of the individual emitters. The correlation achieves a minimum value at zero time delay which given by $g^{(2)}_{off}(0)$ =0.65±0.01, which is close to the ideal value of 0.5. The deviation from the ideal value is due to background emission from the sample and Raman sideband of other detuned quantum dots. In contrast, when we match the frequency of two dots with thermal tuning, we observe an increase in the second-order correlation at zero time delay which is now $g^{(2)}_{on}(0)$ =1.01±0.02, providing a clear signature of super-radiant emission. The width of a sharp peak in the second-order correlation represents a coherence time of the emitters, which we determine to be 140 ps from the numerical fit. This value is consistent with previously measured coherence times.[23] We define the quantum interference visibility as $V = (g_{on}^{(2)}(0) - g_{off}^{(2)}(0))/g_{off}^{(2)}(0)$, which is 1 for ideal case. The measured visibility is 0.55, and the main reason of non-ideal visibility is due to the background emission and limited timing resolution of the detectors of 200 ps, which is similar to the coherence time.

The short coherence time of the emission is a result of the excitation method we use, where we pump the dots with a high-energy laser to create free carriers that naturally diffuse to the ground state of the dot. The timing jitter introduced by this diffusion process degrades the coherence time of the two-emitter system.[29] Resonant or quasi-resonant pumping can eliminate this dephasing mechanism leading to longer coherence times.[30,31]

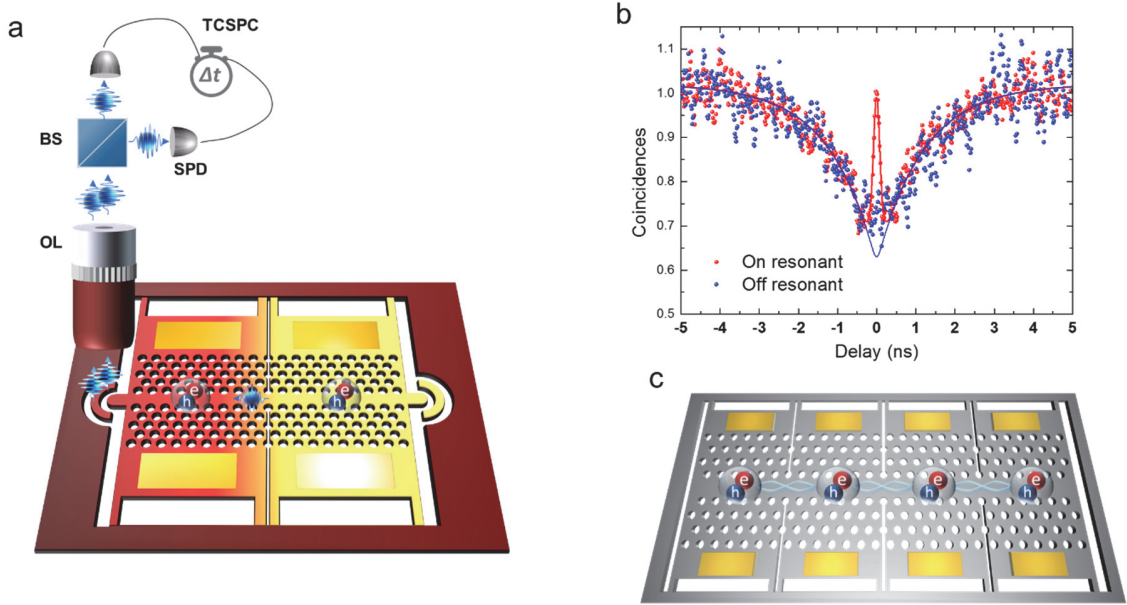

**Figure 3.** (a) Schematic representation of second-order correlation measurement for two quantum emitters on a waveguide. OL: objective lens, BS: 50:50 beam splitter, SPD: single photon detector, TCSPC: time-correlated single photon counter. (b) Second-order photon-correlations histograms when two emitters are on resonant (red dots) and off-resonant (blue dots), respectively. Solid lines are fitted curves. (c) Illustration of interacting multiple identical quantum dots on the controllable waveguide device.

In summary, we have demonstrated the super-radiance from individual quantum dots integrated into a single chip. We resolved the difficulty of long-range interaction and the problem of spectral mismatch between two quantum dots by integrating them onto locally tunable waveguide devices, enabling us to match their frequency and efficiently couple the quantum emitters to the same mode.

Our approach is compatible with other tuning methods, such as strain[32] or quantum confined Stark effect.[33] These methods have the advantage of enabling much wider spectral tuning range without additional linewidth broadening that occurs at higher temperatures for thermal tuning method. Our method also provides a natural path to scale to more quantum dots in the waveguide as illustrated in Figure 3c. Such architectures could enable complex multi-atomic systems with tunable long-range interactions, which could play an important role in scalable and integrated quantum devices for future quantum computation and quantum network.

**Methods.**

*Sample information.* The sample contains InAs/InP quantum dots with a density of approximately 10 $\mu m^{-2}$ in a 280 nm-thick InP membrane on a 2 $\mu m$-thick AlInAs sacrificial layer. The waveguide design was based on a single line defect with a lattice parameter (denoted as *a*) of 350 nm, a hole radius of 0.28 *a*, and a slab thickness of 280 nm. To fabricate the device we deposited a 100 nm-thick silicon nitride thin film as an etching mask using plasma-enhanced chemical vapor deposition. We then patterned the mask using electron beam lithography followed by fluorine-based reactive ion etching and transferred the pattern to the quantum dot sample using chlorine-based reactive ion etching. Finally, we removed the sacrificial layer by selective wet etching to form an air-suspended photonic crystal membrane. We thermally isolated the waveguide devices from the substrate by 500 nm-wide tethers, and also added 150 nm-wide gaps at the middle of photonic crystal devices to thermally separate the waveguides into two regions, left and right. We formed metal heating pads by depositing 5 nm-thick Cr and 35 nm-thick Au on a heating pad for local thermal tuning. The holes at the center of the metal pads were made for clear undercut of the devices during the wet etching process.

*Collection efficiency.* We fabricated the grating out-coupler with a $\lambda/2n$ pitch to extract the waveguide-coupled emission, where $n$ is the refractive index of InP. In the simulation, we calculated that the grating out-coupler could extract about 60% of light from the waveguide toward top direction, while we measured out-coupling efficiency of $\sim$ 19% due to imperfect fabrication and a finite NA=0.7. The waveguide-coupled dots A and B had a collection efficiency of 5% and 4% at the first lens from the left grating out-coupler. We obtained these values from the measured single photon count rates of 4.3 kcounts/s and 2.2 kcounts/s for dots A and B, respectively, at a 5 MHz pulse excitation. Our system has an efficiency about 1.7%, which includes the transmission efficiency of optics (40%) and spectrometer (40%), the coupling efficiency to the fiber (53%), and the detector quantum efficiency (20%).

*Experimental setup.* We mounted the sample on a low-vibration closed cycle cryostat at 4 K. A microscope objective lens (NA=0.7) focused two lasers on two locations for dots A and B and one laser on a heating pad of dot B, and then collected the photons from the left-side grating out-coupler of the devices. For pulsed $g^{(2)}(t)$ and lifetime measurements, we used a 780 nm pulsed laser diode with a pulse width of 50 ps and a repetition rate of 40 MHz and InGaAs single photon detectors with an efficiency of 20%, timing resolution of 200 ps, and dark count rate of 200 Hz. A time-correlated single-photon counting module recorded the histogram for time-resolved lifetime and correlation measurements. To measure the photon-correlation, we coupled the spectrally filtered emission into a fiber-based 50:50 beamsplitter and sent the photons to two InGaAs single photon detectors. We matched the intensity of two quantum dot emissions by changing the laser power. We used about 50% of the saturation power of dots A and B for the photon-correlation measurements.

AUTHOR INFORMATION

**Corresponding Author**
*E-mail: jehyungkim@unist.ac.kr

*E-mail: edowaks@umd.edu

# Supporting Information

## 1. Optical modes in photonic crystal waveguide devices

Figure S1a shows a numerically calculated band structure of an infinitely-long photonic crystal waveguide. The line defect in the photonic crystal structure forms the fundamental (even) and the first higher-order (odd) modes below a light line in the TE-like photonic band gap. The dispersion curves become flat at large $k$ values, resulting in a divergence of photonic density of. Figure S1b exhibits the mode profiles (Ey) of even mode at the wavelength of 1345 nm. When the quantum dots are coupled to this mode, we expect enhanced spontaneous emission rates by Purcell effect. Figure S1c displays the spatial map of simulated Purcell factor ($F_P = \frac{3}{4\pi^2}\left(\frac{\lambda}{n}\right)^3\frac{Q}{V}\left|\boldsymbol{E_y}(\mathbf{r})\right|^2/\left|\boldsymbol{E_{y\,max}}\right|^2$) with a y-polarized dipole, where $Q$, $V$, $\boldsymbol{E_y}(\mathbf{r})$ and $\boldsymbol{E_{y\,max}}$ denote a quality factor, mode volume, electric field at the position of quantum dots, and maximum electric field. We calculate high Purcell factor of 44 at the maximum electric field.

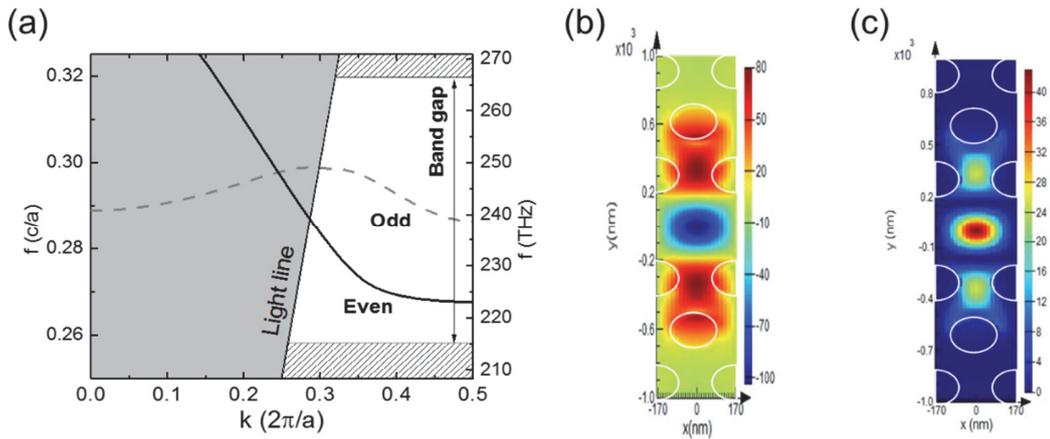

**Figure S1.** (a) Band structure of an infinitely long line-defect waveguide. (b) Electric field (Ey) profiles for even mode of the photonic crystal waveguide. (c) Simulated spatial map of Purcell factor for the even mode with a y-polarized dipole.

## 2. Optical property of quantum dots coupled to photonic crystal waveguides

The slowly varying dispersion curve of the waveguide mode near the band edge has a strong influence on the optical property of quantum dots. In order to characterize the waveguide device, we perform photoluminescence measurements by exciting the center of the waveguide using a 780 nm continuous wave laser and collect the signal at the left grating out-coupler. Figure S2a shows photoluminescence spectra from the devices at low and high excitation powers. At a low excitation power of 200 nW, we observe waveguide-coupled quantum dot emissions over a broad spectral range below 1335 nm, corresponding to the band edge of the fundamental waveguide mode. By increasing the excitation power to 10 μW, we saturate the quantum dot emission and broaden it into a continuum that serves as an internal white light source. This internal source maps out the photonic modes of the waveguide and shows a periodic oscillation due to Fabry-Perot interference induced by reflections from the two ends of the waveguide. From the oscillation period ($\Delta\lambda$) we can estimate a group velocity ($v_g = c/n_g$) of light, where $n_g = \lambda^2/(2l\Delta\lambda)$ is a group index, and $l = 15$ μm is a length of a waveguide.[1] Figure S2b shows group indices at different wavelengths measured from the oscillation period ($\Delta\lambda$). We observe a group index of up to 13 near the band edge of the photonic crystal waveguide.

The waveguide-coupled quantum dots will exhibit Purcell effect. Figure S2c shows the decay curves of the quantum dots A and B in the manuscript. For comparison, we also display the decay curves of the spectrally uncoupled quantum dot on a waveguide and a bulk quantum dot. Compared to the averaged decay rate of bulk quantum dots ($\Gamma_{bulk} \sim 0.45$). The coupled dots A and B show enhanced decay rates ($\Gamma_A$=0.71, $\Gamma_B$=0.68), while the uncoupled quantum dot show a suppressed decay rate ($\Gamma_{uc}$=0.12). In order to confirm Purcell effect, we investigate the decay rates of 33 different quantum dots in 5 different waveguides as shown in Figure S2d and observe a strong enhancement of the decay rate near the band edge, where the photonic density of state becomes maximize. Based on the decay rates ($\Gamma_c$) of these quantum dots, we measure Purcell enhancement ($F = \Gamma_c/\Gamma_{bulk}$) and coupling efficiency ($\beta = (\Gamma_c - \Gamma_{uc})/\Gamma_c$) in the manuscript, where $\Gamma_{bulk}$ denotes the averaged decay rate of the bulk quantum dots.[2] Figure S2e plots the Purcell factor and coupling efficiency of the quantum dots at different wavelengths. Near the band edge, the quantum dots show the Purcell factor as high as 13 and the coupling efficiency as high as 0.98. The measured Purcell factors for the coupled quantum dots much smaller than the simulated value in Figure S1c. We attribute this to imperfect fabrications as well as imperfect spatial and polarization matching between quantum dots and mode. We note that although quantum dots very close to the band edge exhibit high Purcell factors exceeding 10, very little of their emission couples out of the grating, as shown in Figure S2a. We attribute this weak signal to high waveguide losses near the band edge due to scatter[3] and Anderson localization,[4] which is a well-known effect in waveguides with slow group velocities. We find it experimentally favorable to work at a slightly shorter wavelength of 1320 nm, where quantum dots has a moderate Purcell factor of 2~3, but they still show a high coupling efficiency of 80%, and the waveguides are far less lossy.

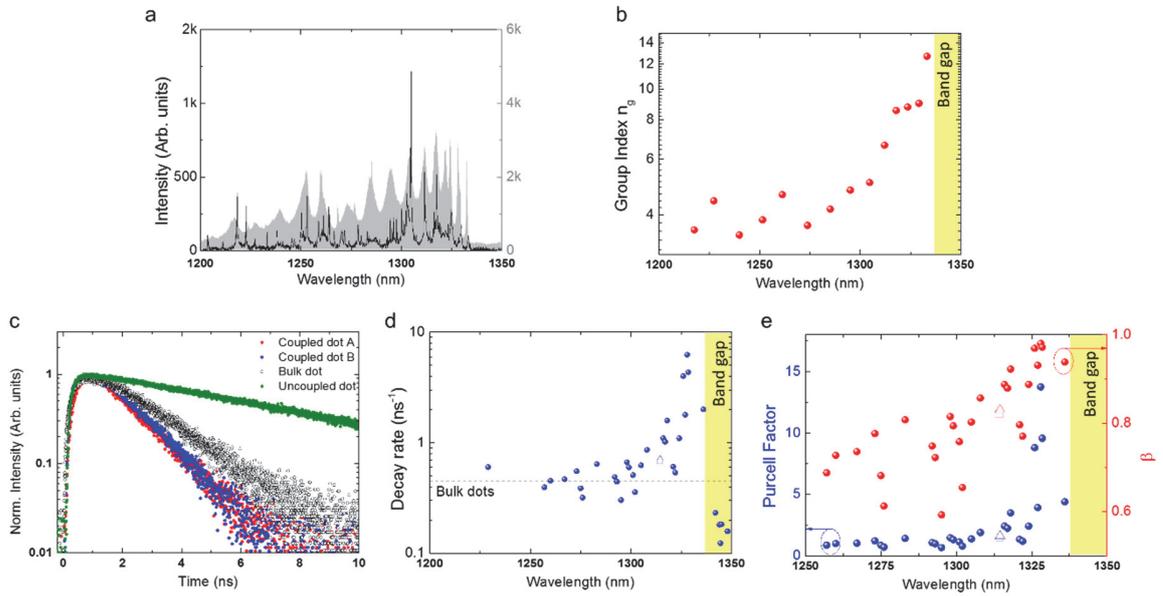

**Figure S2.** (a) Photoluminescence spectra with low excitation (black) and high excitation (gray) powers that show waveguide-coupled quantum dot emissions and waveguide modes, respectively. (b) Measured group indices in the waveguide at different wavelengths. (c) Comparison of decay curves of the waveguide-coupled quantum dots A (red solid) and B (blue solid), uncoupled quantum dots on the waveguide (green solid), and the quantum dot in a bulk region (black empty). (d) Measured decay rates of several quantum dots on waveguides at different wavelengths. Dashed line denotes an averaged decay rate of quantum dots in InP bulk. Empty blue triangle and square at about 1314 nm represent the decay rates of the studied quantum dots A and B in the manuscript. (e) Purcell factor (blue solid-circles) and coupling efficiency (red solid-circles) of the quantum dots on waveguides at different wavelengths. Empty triangles and squares at about 1314 nm represent the Purcell factor and coupling efficiency of the studied quantum dots A and B.

## 3. Temperature dependence of single quantum dot emission

We tune the quantum dot wavelength using a local heating technique. However, increasing the temperature not only changes the wavelength but also affects the intensity and linewidth of the quantum dots due to enhanced interaction with phonons. Figure S3b-d shows a change of the wavelength, intensity and linewidth of the quantum dot emission as a function of temperature with a sample stage heating. As we change the temperature from 4 K to 36 K, the wavelength red shifts about 3 nm. From these results, we determine the local quantum dot temperature of about 15 K when we locally heat the quantum dot B with a laser in Figure 2e. At this temperature the intensity and line width of the quantum dots are almost same as the low temperature at 4 K, while they start to reduce and broaden significantly above 26 K.

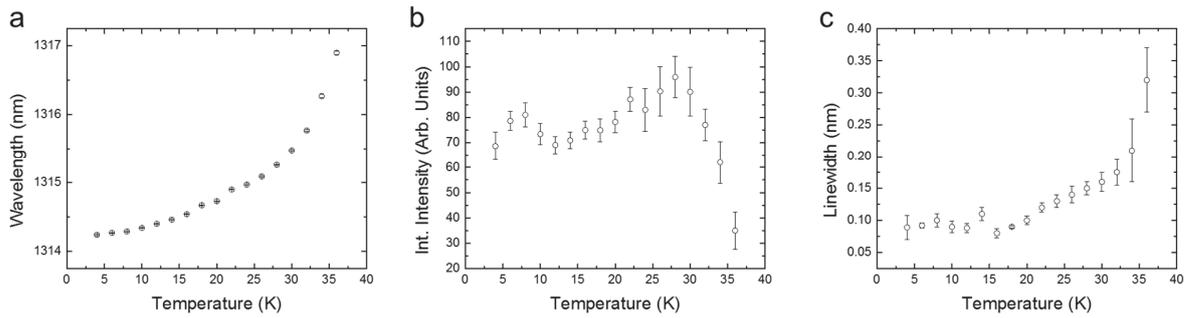

**Figure S3.** (a-c) Wavelength (a), integrated intensity (b), and linewidth (c) of the quantum dot emission as a function of stage heating temperature.

## 4. Theoretical model for two quantum dots in a waveguide

Figure S4 illustrates the model we analyze. Two emitters with dipole operators $\sigma_-$ and $\mathbf{s}_-$ are coupled to a waveguide whose output operator is denoted $\mathbf{a}$. The dipole operators and waveguide operator are related by the input-output relation

$$\mathbf{a} = \sqrt{\gamma_{wg}}\left(\sigma_- + e^{ikL}\mathbf{s}_-\right) \tag{1}$$

where $\gamma_{wg}$ is the radiative coupling rate of the emitters into the waveguide mode.

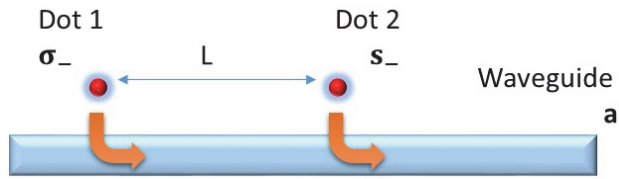

**Figure S4.** Two emitters separated by L with dipole operators $\sigma_-$ and $\mathbf{s}_-$ are coupled to the waveguide whose output operator is $\mathbf{a}$.

For continuous wave excitation the system is described by a stationary process, and the un-normalized second-order correlation function for the waveguide mode is given by

$$G^{(2)}(\tau) = \left\langle \mathbf{a}^\dagger(0)\mathbf{a}^\dagger(\tau)\mathbf{a}(\tau)\mathbf{a}(0) \right\rangle \tag{2}$$

Using the input-output relations in Eq. (1), we obtain

$$G^{(2)}(\tau) = G_a^{(2)}(\tau) + G_b^{(2)}(\tau) + 2G_a^{(1)}(0)G_b^{(1)}(0) + G_a^{(1)}(\tau)R_b^{(1)}(\tau) + G_b^{(1)}(\tau)R_a^{(1)}(\tau) \tag{3}$$

where

$$
\begin{aligned}
G_a^{(2)}(\tau) &= \gamma_{wg}^2 \left\langle \sigma_+(0)\sigma_+(\tau)\sigma_-(\tau)\sigma_-(0) \right\rangle \\
G_b^{(2)}(\tau) &= \gamma_{wg}^2 \left\langle \mathbf{s}_+(0)\mathbf{s}_+(\tau)\mathbf{s}_-(\tau)\mathbf{s}_-(0) \right\rangle \\
G_a^{(1)}(\tau) &= \gamma_{wg} \left\langle \sigma_+(0)\sigma_-(\tau) \right\rangle \\
G_b^{(1)}(\tau) &= \gamma_{wg} \left\langle \mathbf{s}_+(0)\mathbf{s}_-(\tau) \right\rangle \\
R_a^{(1)}(\tau) &= \gamma_{wg} \left\langle \sigma_+(\tau)\sigma_-(0) \right\rangle \\
R_b^{(1)}(\tau) &= \gamma_{wg} \left\langle \mathbf{s}_+(\tau)\mathbf{s}_-(0) \right\rangle
\end{aligned}
\tag{4}
$$

The first-order correlation functions satisfy the differential equations

$$\frac{dG_{a,b}^{(1)}}{d\tau} = -\left(i\Delta_{a,b} + \beta\right)G_{a,b}^{(1)}$$
$$\frac{dR_{a,b}^{(1)}}{d\tau} = \left(i\Delta_{a,b} - \beta\right)R_{a,b}^{(1)}$$

(5)

In the above equations $\beta = \frac{\gamma}{2} + \frac{1}{T_2}$ where $\gamma$ is the radiative decay rate of the emitter and $T_2$ is the dipole dephasing time. For simplicity we assume the two emitters have identical decay rates and dephasing times. $\Delta_{a,b}$ represents the frequency detuning of the two emitters from the optical mode. The solution to the above equations is given by

$$G_{a,b}^{(1)}(\tau) = G_{a,b}^{(1)}(0)e^{-\left(i\Delta_{a,b} + \beta\right)\tau}$$
$$R_{a,b}^{(1)}(\tau) = G_{a,b}^{(1)}(0)e^{\left(i\Delta_{a,b} - \beta\right)\tau}$$

(6)

Plugging these results into Eq. (3) we obtain

$$G^{(2)}(\tau) = G_a^{(2)}(\tau) + G_b^{(2)}(\tau) + 2G_a^{(1)}(0)G_b^{(1)}(0) + 2\cos(\Delta\tau)e^{-2\beta\tau}G_a^{(1)}(0)G_b^{(1)}(0)$$

(7)

where $\Delta = \Delta_a - \Delta_b$ is the detuning between the two emitters.

The normalized second order correlation is given by

$$g^{(2)}(\tau) = \frac{\left\langle \mathbf{a}^\dagger(0)\mathbf{a}^\dagger(\tau)\mathbf{a}(\tau)\mathbf{a}(0) \right\rangle}{\left\langle \mathbf{a}^\dagger(0)\mathbf{a}(0) \right\rangle \left\langle \mathbf{a}^\dagger(\tau)\mathbf{a}(\tau) \right\rangle}$$

(8)

If we assume both emitters have the same radiative decay rate into the waveguide then

$$\left\langle \mathbf{a}^\dagger(0)\mathbf{a}(0) \right\rangle = \left\langle \mathbf{a}^\dagger(\tau)\mathbf{a}(\tau) \right\rangle = 2G_a^{(1)}(0)$$

(9)

and

$$g^{(2)}(\tau) = \frac{1}{2}\left(g_a^{(2)}(\tau) + 1\right) + \frac{\cos(\Delta\tau)}{2}e^{-\beta\tau}$$

(10)

The above expression is a sum of two terms. The first term is identical to the second-order correlation function for two distinguishable emitters, while the second is an interference term that depends on the detuning and coherence of the two emitters.

For an ideal single photon source, $g_a^{(2)}(\tau) = 1 - e^{-\gamma t}$ which leads to the final expression

$$g^{(2)}(\tau) = \frac{1}{2}\left(2 + \cos(\Delta \tau)e^{-\beta \tau} - e^{-\gamma t}\right) \qquad (11)$$

We consider two special cases of the above equation: (1) Distinguishable emitters, and (2) Indistinguishable emitters.

**Distinguishable emitters**: We consider the case of two distinguishable emitters that emit at different frequencies. Specifically, we consider the case where $\delta \tau \gg 1/\Delta$ where $\delta \tau$ is the time resolution of the measurement system. In our case this time resolution is limited by the photon counters to approximately 200 ps. When the above condition holds, the cosine term in Eq. (11) oscillates quickly compared to the time response of the system and rapidly averages to zero. The second-order correlation becomes

$$g^{(2)}(\tau) = \frac{1}{2}\left(2 - e^{-\gamma t}\right) \qquad (12)$$

which is the expected result for two distinguishable emitters coupled to a single mode. At zero time delay we attain $g^{(2)}(0) = \frac{1}{2}$, providing the expected anti-bunching behavior of a two-emitter system.

**Indistinguishable emitters**: In this case $\Delta = 0$ and the expression in Eq. (11) becomes

$$g^{(2)}(\tau) = \frac{1}{2}\left(2 + e^{-\beta\tau} - e^{-\gamma t}\right) \tag{13}$$

The second-order correlation at zero delay now takes on the value $g^{(2)}(0) = 1$, which is twice the value of the case for distinguishable emitters.